\documentclass[12pt]{iopart}

\usepackage{psfrag,graphicx}
\usepackage{amsfonts,amssymb,amsmath}        

\newcommand{\1}{1\!\!1}

\newcommand{\ket}[1]{\left | \, #1 \right\rangle}
\newcommand{\bra}[1]{\left \langle #1 \, \right |}

\newcommand{\rf}[1]{(\ref{#1})}

\begin{document}

\title{Non-Abelian statistics as a Berry phase in exactly solvable models}

\author{Ville Lahtinen and Jiannis K Pachos}
\address{School of Physics \& Astronomy, University of Leeds,
Leeds LS2 9JT, UK }
\date{\today}
\ead{pyvtl@leeds.ac.uk (Ville Lahtinen)}

\begin{abstract}

We demonstrate how to directly study non-Abelian statistics for a wide class of exactly solvable many-body quantum systems. By employing exact eigenstates to
simulate the adiabatic transport of a model's quasiparticles, the
resulting Berry phase provides a direct demonstration of their non-Abelian
statistics. We apply this technique to Kitaev's honeycomb lattice model and explicitly demonstrate the existence of non-Abelian Ising anyons confirming the previous conjectures. Finally, we present the manipulations needed to transport and
detect the statistics of these quasiparticles in the laboratory. Various physically realistic system sizes are considered and exact predictions for such experiments are provided.

\end{abstract}

\pacs{05.30.Pr, 75.10.Jm}

\maketitle

\section{Introduction}

A striking feature of topological phases of matter is that they can support
anyons. These are quasiparticles with statistics different from bosons or fermions. The statistical behavior of
anyons is demonstrated by their adiabatic exchange, which causes non-trivial
evolution in their quantum state. For Abelian anyons the evolution is given by a phase factor. The presence of non-Abelian anyons gives rise to degeneracy in the energy spectrum and the evolution is
described by a unitary matrix acting on the degenerate states. In general, anyons are known to exist in different varieties distinguished by their characteristic statistics \cite{Rowell}. Therefore, the explicit demonstration of the statistical behavior is essential for the unique characterization of a topological phase. Such
phases are of great interest due to the possibility of realizing anyons in a physical system
and due to their potential for technological applications. In particular,
topological quantum computation employs the anyonic statistics for
performing error-free quantum information processing~\cite{Freedman04}.

The best known many-body system conjectured to support non-Abelian
statistics is the fractional quantum Hall
liquid~\cite{Moore, Read96, Read00}. Other proposals include the $p$-wave
superconductor~\cite{Read00, Gurarie} as well as various lattice
models~\cite{Doucot,Freedman05,Kitaev05}.
These systems are either tailored to identically
support non-Abelian statistics and have complex physical realizations, or
they can be described by simple Hamiltonians, but their statistical behavior
is based on indirect arguments. In particular, for the fractional quantum Hall states
they rely on properties of trial wave functions~\cite{Arovas,Tserkovnyak,Read08}, whereas for the lattice models explicit calculations have not been previously attempted. Although the indirect arguments are sound, direct calculations of the statistics are crucial to resolve any ambiguities, to address physical realizable finite-size systems and to provide exact predictions for the experiments.

Here we demonstrate how to directly calculate the non-Abelian
statistics for a class of exactly solvable models. By applying the Berry phase technique \cite{Arovas} to the Kitaev's honeycomb spin lattice model~\cite{Kitaev05}, we calculate the
evolution associated with an adiabatic exchange of quasiparticles. This is performed using exact eigenstates of a 360 spin system. We obtain a unitary matrix that corresponds to the
statistics of the conjectured non-Abelian Ising anyons. Together with the
fusion rules of these anyons~\cite{Lahtinen,Pachos06}, this conclusively
demonstrates the non-Abelian character of Kitaev's model, thereby confirming the conjectured behavior. Further, we
present a scheme for creating, transporting and characterizing the anyons
that could be used in the proposed physical
implementations~\cite{Micheli05} and provide exact predictions for a physically realistic range of the model's parameters.

\section{The honeycomb lattice model}

\begin{figure}[t]
\begin{indented}
\item[]\includegraphics*[width=9cm]{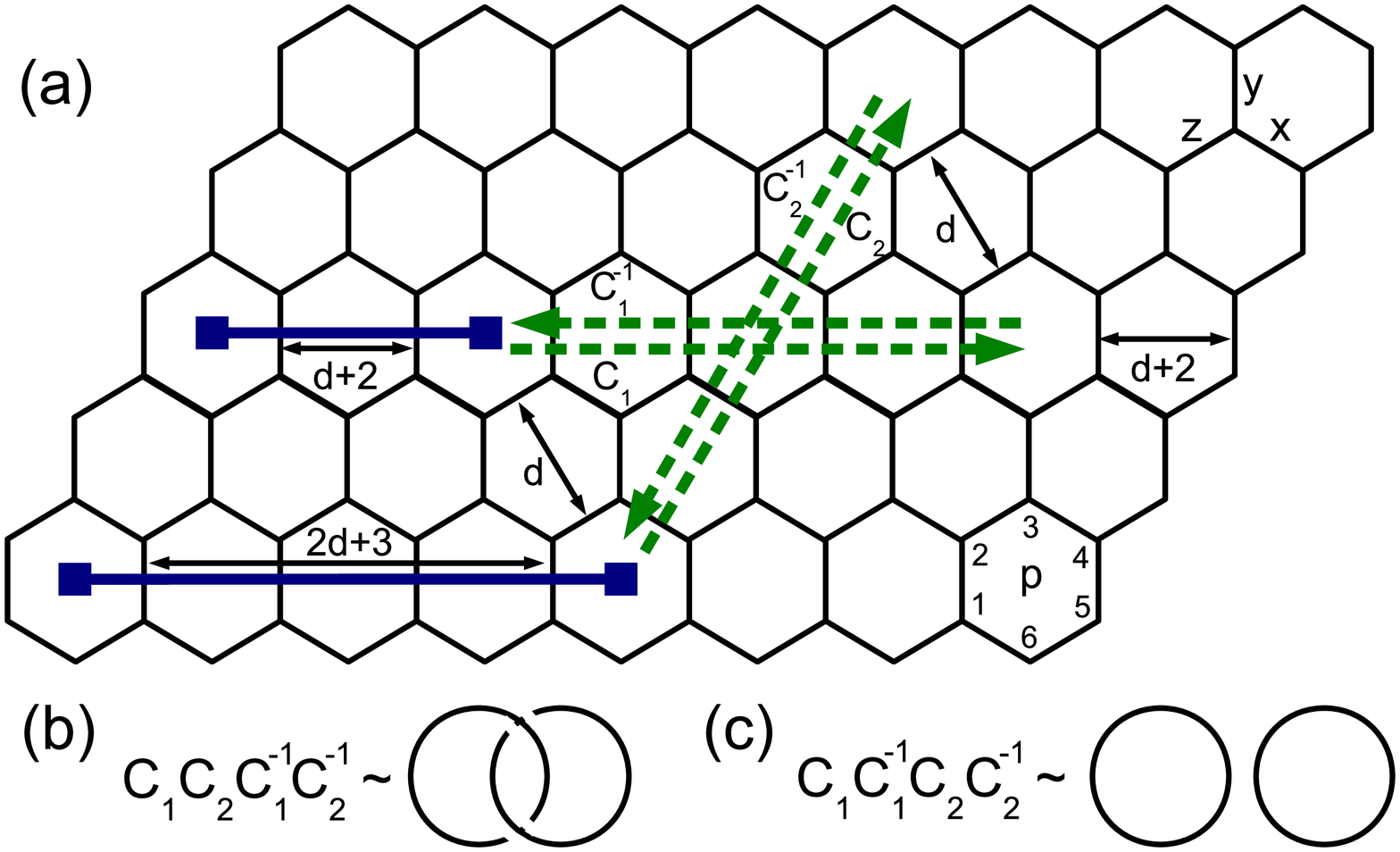}
\end{indented}
\caption{\label{honeycomb} (a) The honeycomb lattice on a torus
containing two vortex pairs. This vortex configuration is created by setting
$u_{ij} = -1$ on the links crossed by solid lines and $u_{ij} = 1$ on all
other links. The parameter $d$ controls the minimal vortex separation. It is
related to the torus dimensions through $M=2(2d+4)$ and $N=2d+3$ (picture not
on scale). The four dashed arrows $C_1$, $C_1^{-1}$, $C_2$ and $C_2^{-1}$
are the oriented parts of the path $C$ along which the vortices are moved.
(b) $C_{l}=C_1 C_2 C_1^{-1} C_2^{-1}$ is topologically equivalent to a link.
(c) $C_{o}=C_1 C_1^{-1} C_2 C_2^{-1}$ is topologically equivalent to two
unlinked loops.}
\end{figure}

Kitaev's model~\cite{Kitaev05} comprises of spin-$1/2$ particles residing
on the vertices of a honeycomb lattice. The spins interact according to the
Hamiltonian
\begin{eqnarray} \label{H}
    H = - \sum_{\nu \in \{x,y,z\}} \sum_{(i,j) \in \nu \textrm{-links}} J_{ij}^\nu \sigma^\nu_i \sigma^\nu_j -
    \sum_{( i,j,k)} K_{ijk} \sigma^x_i \sigma^y_j \sigma^z_k,
\end{eqnarray}
where $J_{ij}^\nu$ are positive nearest neighbor couplings on links $(i,j)$ of type $\nu$ (see \Fref{honeycomb}(a) for link labeling).
The second term is an effective magnetic field
with positive next-to-nearest neighbor couplings $K_{ijk}$, such that every plaquette $p$ contributes the six terms
\begin{eqnarray}
\sum_{( i,j,k) \in p} K_{ijk} \sigma^x_i \sigma^y_j \sigma^z_k & = & K_{123}\sigma^z_1 \sigma^y_2 \sigma^x_3+K_{234}\sigma^x_2 \sigma^z_3
\sigma^y_4+K_{345}\sigma^y_3 \sigma^x_4 \sigma^z_5+ \nonumber \\ 
\ & \ & K_{456}\sigma^z_4 \sigma^y_5 \sigma^x_6+K_{561}\sigma^x_5 \sigma^z_6 \sigma^y_1+K_{612}\sigma^y_6 \sigma^x_1 \nonumber
\sigma^z_2.
\end{eqnarray}
The enumeration of the sites is shown in \Fref{honeycomb}(a). The Hamiltonian has the
symmetry $[H,\hat{w}_p]=0$, where $\hat{w}_p =
\sigma^x_1 \sigma^y_2
\sigma^z_3
\sigma^x_4 \sigma^y_5 \sigma^z_6$ are plaquette operators whose
eigenvalues $w_p = - 1$ are interpreted as having a {\em vortex} on
plaquette $p$. We represent the spin operators as
$\sigma^{\nu}_i = i b^\nu_i c_i$, where $c_i, b^x_i, b^y_i$ and
$b^z_i$ are Majorana fermions \cite{Kitaev05, Lahtinen}. Subsequently, the Hamiltonian takes the form
$H = \frac{i}{4} \sum_{i,j} \hat{A}_{ij} c_{i} c_{j}$, where
\begin{eqnarray} \label{A}
    \hat{A}_{ij} = 2J_{ij} \hat{u}_{ij} + 2 \sum_{k} K_{ijk} \hat{u}_{ik}
    \hat{u}_{jk}, \quad \hat{u}_{ij} = ib^\nu_i b^\nu_j.
\end{eqnarray}
Here $J_{ij}$ and $\hat{u}_{ij}$ are shorthand notations for $J^\nu_{ij}$ and $\hat{u}_{ij}^\nu$ when $(i,j)$ is an $\nu$-link.
Since the mapping to Majorana fermions doubles the size of the Hilbert space, the eigenstates of the original Hamiltonian \rf{H} are subject to the constraint
\begin{equation} \label{D}
D_i \ket{\Psi} = \ket{\Psi}, \quad D = b^x_i b^y_i b^z_i c_i, \quad
[D_i,\sigma_j^\nu]=0,
\end{equation}
which follows from the operator identity $\sigma_i^x \sigma^y_i \sigma^z_i = b^x_i b^y_i b^z_i c_i = \1$. Since  $[H, \hat{u}_{ij}] = 0$, the Hilbert space splits into sectors each
labelled by $u$, a certain pattern of eigenvalues $u_{ij}$. The
configurations $u$ can be understood as a classical $Z_2$ gauge field with
local gauge transformation operators $D_i$. Consequently, the plaquette operators
$\hat{w}_p =\prod_{(i,j) \in p} \hat{u}_{ij}$ can be identified with gauge
invariant Wilson loop operators, whose patterns of eigenvalues label the physical sectors of the model. Fixing the gauge field configuration $u$
gives then a particular vortex configuration. Throughout this paper we use $J^\nu, K$ and $u$ without indices to denote global configurations of these local quantities and use indices only when referring to their local values. For instance, $K = a$ means that $K_{ijk} = a$ for all $i$, $j$ and $k$.

Consider the system of $2MN$ spins on a torus and assume
$u$ to be fixed such that it creates the four vortex
configuration shown in~\Fref{honeycomb}(a). Diagonalization reduces the Hamiltonian to the canonical form $H~=~\sum_{k = 1}^{MN}  \epsilon_{k} [ b_{k}^{\dagger} b^{ \ }_k - \frac{1}{2} ]$,
where $b_{k}$ are fermionic operators satisfying $\{
b_{k}^\dagger, b^{ \ }_{l}
\} =\delta_{kl}$ and $\epsilon_{k}$ are the corresponding positive eigenvalues. In \cite{Lahtinen} it was shown that when the system is in the
non-Abelian phase ($J_x = J_y = J_z$ = 1, $K>0$), the presence of $2n$ {\em well separated} vortices gives rise to $n$ {\em zero modes} ($\epsilon_{k} \approx 0$, for
$k=1,\ldots,n$). Importantly, these are separated from the rest of the
fermionic spectrum by a finite energy gap. In our case of four vortices this
implies fourfold ground state degeneracy arising from a pair of zero modes
that can be either occupied or empty
\begin{eqnarray} \label{GSs}
    \ket{\Psi_{\alpha_1 \alpha_2}} & = & (b^\dagger_1)^{\alpha_1} (b^\dagger_2)^{\alpha_2}
    \ket{\textrm{gs}},
\end{eqnarray}
where $\alpha_1,\alpha_2=0,1$ and $\ket{\textrm{gs}}= \prod_{k=1}^{MN}
b_k \ket{\phi}$ is the ground state. For convenience we choose
the reference state such that $b_{k}^\dagger \ket{\phi} = 0$.

Numerical diagonalization of $A$, (see \rf{A}), gives $2MN$
eigenvectors $\psi_{k}^\pm$ satisfying the double spectrum $A \psi_{k}^\pm =
\pm \epsilon_{k}\psi_{k}^\pm$, where $\epsilon_{k}$ coincide with the positive eigenvalues of the
diagonalized Hamiltonian.
We construct a representation of the two degenerate ground states
$\ket{\Psi_{10}}$ and $\ket{\Psi_{01}}$ as
\begin{eqnarray} \label{basisrep}
    \ket{\Psi_{\alpha}} =  \sum_{\substack{\{ k,\ldots,l = 1 | \\
    k,\ldots,l \neq \alpha \}}}^{MN-1} \frac{\varepsilon_{k,\ldots,l}}
    {\sqrt{(MN-1)!}}\psi_k^- \otimes \cdots \otimes \psi_l^-,
\end{eqnarray}
where $\alpha=1,2$, respectively, and $\varepsilon_{k,\ldots,l}$ is the fully
anti-symmetric tensor of rank $MN-1$. In general, such states are too large
to be stored in a computer, because their number of elements grows
exponentially with the system size. However, the inner product of two such
vectors, each depending possibly on some parameters $t$ and $t'$, can be efficiently calculated and is given
by
\begin{eqnarray} \label{inner}
    \langle \Psi_{\alpha}(t) \ket{\Psi_{\beta}(t')} = \det ( B_{\alpha \beta}^{tt'} ),
\end{eqnarray}
where $ [B_{\alpha \beta}^{tt'}]_{kl} = \psi_k^{-\dagger}(t) \psi_l^-(t')$.

\subsection{The Ising anyon model}

It has been conjectured that Kitaev's model
supports the Ising anyon model~\cite{Kitaev05,Moore,Read00}. This model has three types
of particles: $1$ (vacuum), $\psi$ (fermion) and $\sigma$ (non-Abelian
anyon). In~\cite{Lahtinen} these are identified with the ground state, the
fermion modes $b^{\dagger}$ and the vortices,
respectively. The non-trivial fusion rules are given by $\psi\times\psi =
1,\quad\psi\times \sigma = \sigma$ and $\sigma \times
\sigma = 1 +\psi$. The last fusion rule implies that there is a degree of freedom
associated with the different ways a number of $\sigma$'s can fuse when
their total anyonic charge is fixed. Taking the four $\sigma$ particles to
fuse to a $\psi$, this fusion degree of freedom is encoded in the two dimensional {\em fusion space}, $V_{\sigma^{4}}^\psi$.
Its basis can be chosen to be the states associated with the
two distinct pair-wise fusion channels:
\begin{equation} \label{Vbasis}
\begin{array}{rcl}
    (\sigma \times \sigma) \times (\sigma \times \sigma) & \to & \psi
    \times 1 = \psi, \\
    (\sigma \times \sigma) \times (\sigma \times \sigma) & \to & 1
    \times \psi = \psi.
\end{array}
\end{equation}
In \cite{Lahtinen} the number of intermediate $\psi$'s is identified with
the number of occupied zero modes. Hence, a suitable basis is given by the
states $ \{\ket{\Psi_{1}}, \ket{\Psi_{2}} \}$ (see \rf{basisrep}). The braid
operator, $R$, describes the statistics of the $\sigma$
 anyons. In particular, the {\em monodromy} operator, $R^2$, corresponds to one
particle encircling another clockwise. On the basis~\rf{Vbasis} the
monodromy of two $\sigma$'s that belong to different pairs is given by
\begin{eqnarray} \label{monodromy}
R^2 = e^{-\frac{\pi}{4}i} \left( \begin{array}{cc} 0 & 1 \\ 1 & 0 \end{array} \right).
\end{eqnarray}

\section{Non-Abelian statistics as a holonomy}

When $z_1$ and $z_2$ are the coordinates of the $\sigma$ anyons, their statistics is given by the transformation of the wave function under their permutation, i.e. $\psi(z_1,z_2) = U \psi(z_2,z_1)$ with $U$ being the characteristic statistical phase or matrix. In real physical systems the permutation of the coordinates corresponds to adiabatically transporting the anyons such that their positions are swapped. When the positions are swapped twice, i.e. a particle winds around the other along a suitable chosen path, the statistics corresponds to the accumulated wave function evolution, which is given by the Berry phase, or the \emph{holonomy} \cite{Arovas, Pachos00}. For bosons and fermions this is always trivial, with non-trivial evolution being a sign of anyonic statistics.

We demonstrate the statistics of $\sigma$ anyons by
adiabatically transporting a vortex around another. Consider a Hamiltonian $H(\lambda)$ with $n$-fold degeneracy
$\{\ket{\Psi_\alpha(\lambda)} | \alpha=1,\ldots,n \}$ that depends on some parameters
$\lambda$. When we adiabatically vary $\lambda$ along a closed path $C$, the
evolution of the degenerate subspace is given by the holonomy $\Gamma_C =
P\exp \oint_C A^\mu(\lambda)d\lambda_\mu$, where $[A^\mu(\lambda)]_{\alpha\beta}
=\bra{\Psi_\alpha(\lambda)}\frac{d}{d\lambda^\mu}
\ket{\Psi_\beta(\lambda)}$ and $P$ denotes path ordering in $\lambda$. To
simulate the vortex transport, we discretize the path $C$ into $T$
infinitesimal intervals of length $\delta\lambda$ with $\lambda(t)$ denoting
the control parameter value at step $t$. It follows that the holonomy takes
the form
\begin{eqnarray} \label{dhol}
    \Gamma_C = \lim_{T \to \infty} P \prod_{t=1}^T \left(\sum_{\alpha=1}^n|\Psi_\alpha\big(\lambda(t)\big)
    \rangle\langle\Psi_\alpha\big(\lambda(t)\big)|\right),
\end{eqnarray}
i.e. in the limit $\delta \lambda \to 0$ it is given by the ordered product of projectors onto the ground state
space at each step $t$.

We evaluate the evolution in the fusion space $V_{\sigma^{4}}^\psi$. The basis states~\rf{basisrep} are not symmetrized under gauge transformations \rf{D}. Nevertheless, their holonomy coincides with the holononomy of symmetrized states when $C$ is a loop in both the space of four vortex and gauge field configurations. This is due to the orthogonality of states belonging to different sectors of $u$. A suitable path is illustrated in~\Fref{honeycomb}(a), where the path $C$ (dashed lines) is split into
four parts. Different ordering of these parts corresponds to the
topologically inequivalent paths $C_{l}$ and $C_{o}$ given in~\Fref{honeycomb}(b) and (c), respectively. Neither path spans any area
and hence all contribution to the holonomy is topological. Since $u$ is a
static background field, we need to introduce classical control parameters
to physically implement the transport. Assuming local control of $J_{ij}$ and $K_{ijk}$
on all links, we see from~\rf{A} that the simultaneous sign change of these quantities on link $(i,j)$ is equivalent to changing $u_{ij} \to -u_{ij}$. This either generates a vortex pair or transports a vortex through the link $(i,j)$. In our simulation this is performed in $S$
infinitesimal steps. Taking $\lambda = (J,K)$ and assuming $T$ to be a sufficiently large, the discrete
holonomy~\rf{dhol} for the degenerate states
\rf{basisrep} is well approximated by
\begin{eqnarray} \label{holonomy}
    \Gamma_C \approx P \prod_{t=1}^{T-1} \left( \begin{array}{cc} \det
    ( B^{t,t+1}_{11} ) & \det ( B^{t,t+1}_{12} ) \\
    \det ( B^{t,t+1}_{21} ) & \det ( B^{t,t+1}_{22} )  \end{array} \right),
\end{eqnarray}
where we have used the inner product \rf{inner}\footnote{The freedom in changing the basis at each step $t$ of the path $C$ gives rise to an accumulated unitary matrix, $M$, making the result of the adiabatic evolution to be in general given by $M\Gamma_C$~\cite{Read08}. Here we choose $\ket{\Psi_\alpha(t=0)}=\ket{\Psi_\alpha(t=T)}$ which gives $M=\1$.}. Therefore, the holonomy can
be evaluated by diagonalizing the Hamiltonian at each step $t$ and
multiplying together the inner products of the eigenstates from successive
steps according to \rf{holonomy}. We perform this for the three
parametrizations shown in Table \ref{parconfs}.
\begin{table}[h]
\caption{ \label{parconfs} Three parametrizations (i), (ii) and (iii)
for which the holonomy is evaluated. Here $T=8S(d+1)$ and the number of
spins is $2MN=8(d+2)(2d+3)$. $S$ has been increased in (iii) to suppress accumulation of discretization errors due to longer path.}
\begin{indented}
\item[]\begin{tabular}{ccccc}
\hline
 \ & $d$ & $S$ & $T$ & $2MN$ \\
\hline
  \ (i) \ & \ 1 \ & \ 2$\cdot 10^3$ \ & 32$\cdot 10^3$ & \ 120 \ \\
\hline
 \ (ii) \ & \ 2 \ & \ 2$\cdot 10^3$ \ & 48$\cdot 10^3$ & \ 224 \ \\
\hline
 \ (iii) \ & \ 3 \ & \ 4$\cdot 10^3$ \ & 128$\cdot 10^3$ & \ 360 \ \\
\hline
\end{tabular}
\end{indented}
\end{table}

\begin{figure}[t]
\begin{indented}
\item[]\includegraphics*[width=9cm]{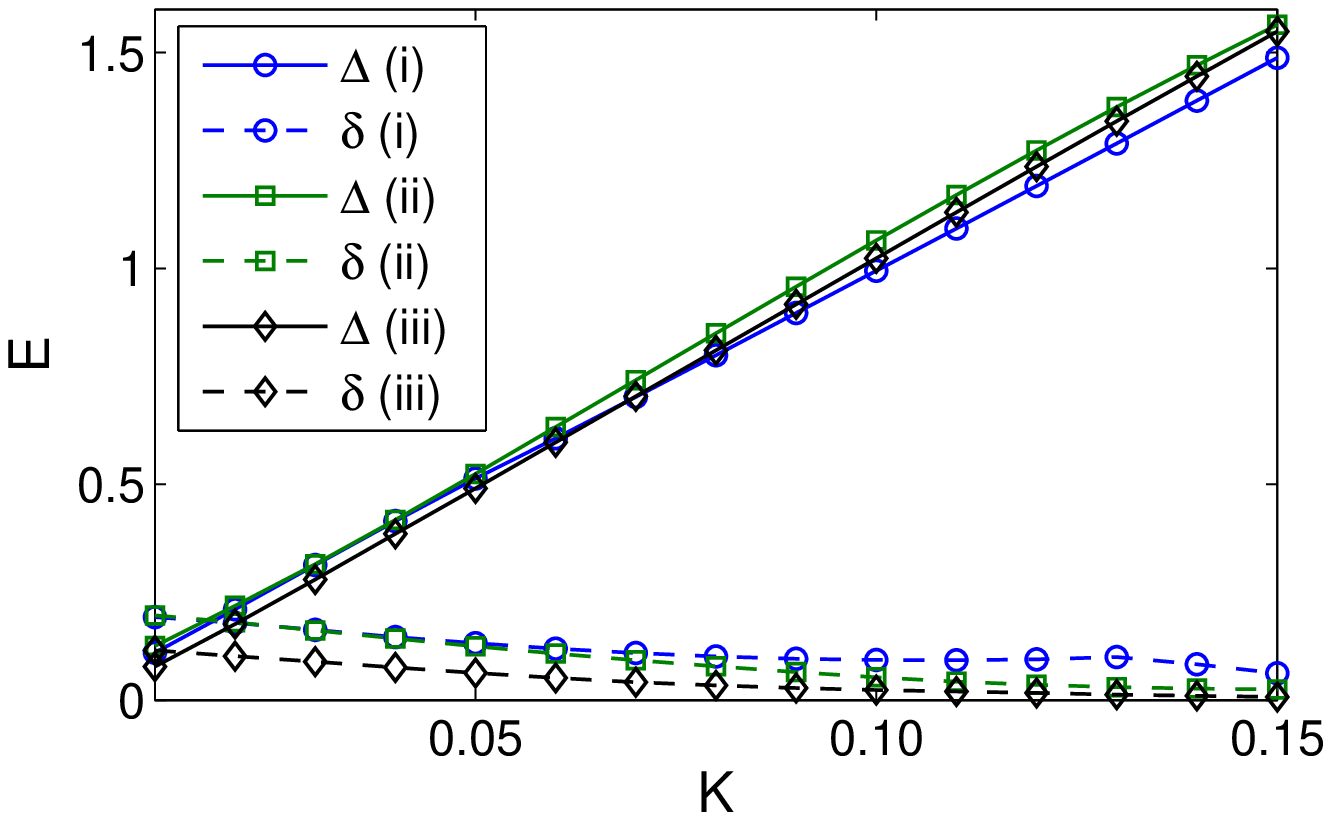}
\end{indented}
\caption{\label{gaps} The minimal fermion gap $\Delta$
(\full) and the maximum energy splitting between the ground states
$\delta$,~\rf{gapdeg} (\dashed) as functions of $K$ for
parametrizations (i) (\opencircle), (ii) (\opensquare) and (iii) (\opendiamond) given in
Table~\ref{parconfs}. The fermion gap grows linearly and the degeneracy
improves with increasing $K$ for all parametrizations. The fermion gap is
relatively insensitive to the vortex separation, whereas the degeneracy
improves when the vortices are further apart.}
\end{figure}

\begin{figure}[t]
\begin{indented}
\item[]\begin{tabular}{c}
\includegraphics*[width=9cm]{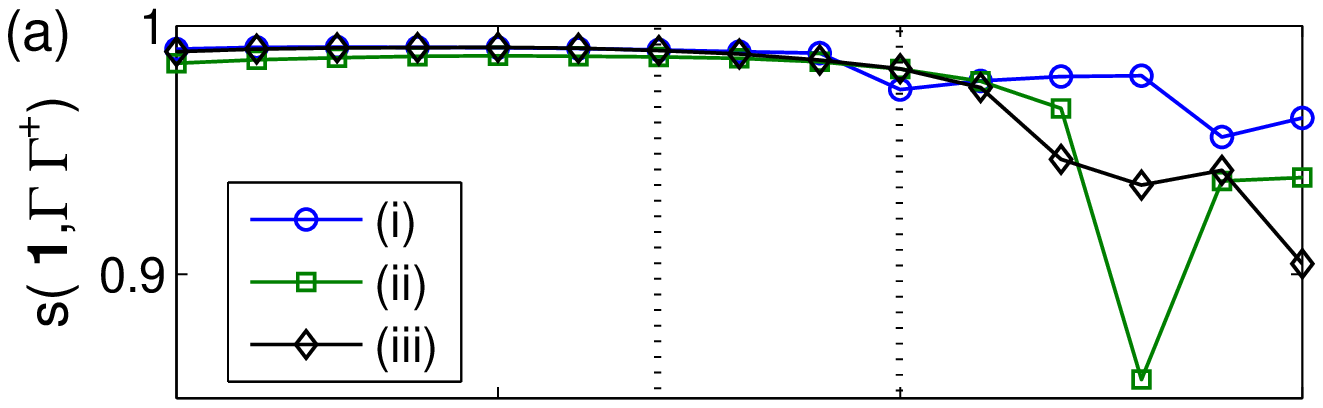} \\
\includegraphics*[width=9cm]{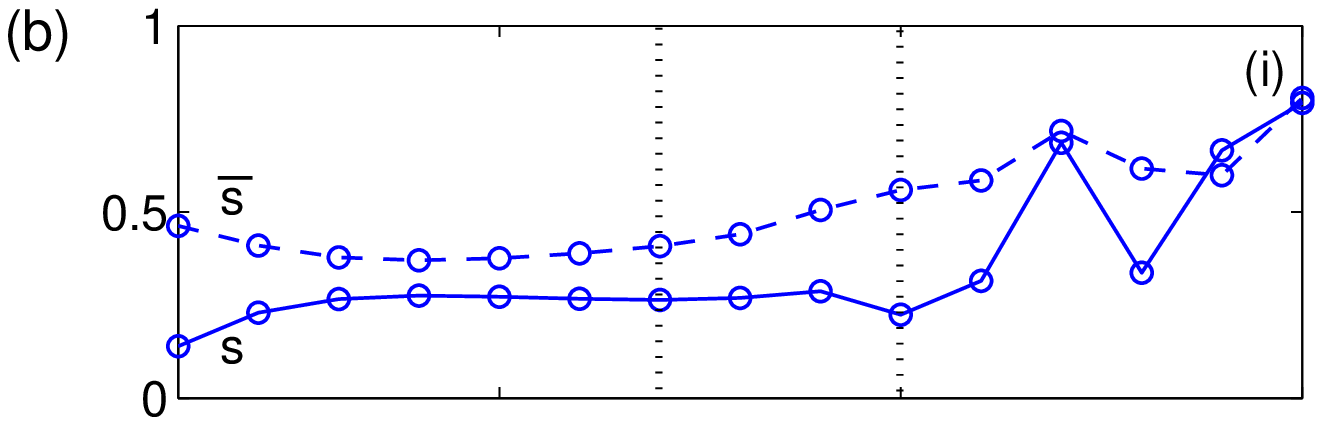} \\
\includegraphics*[width=9cm]{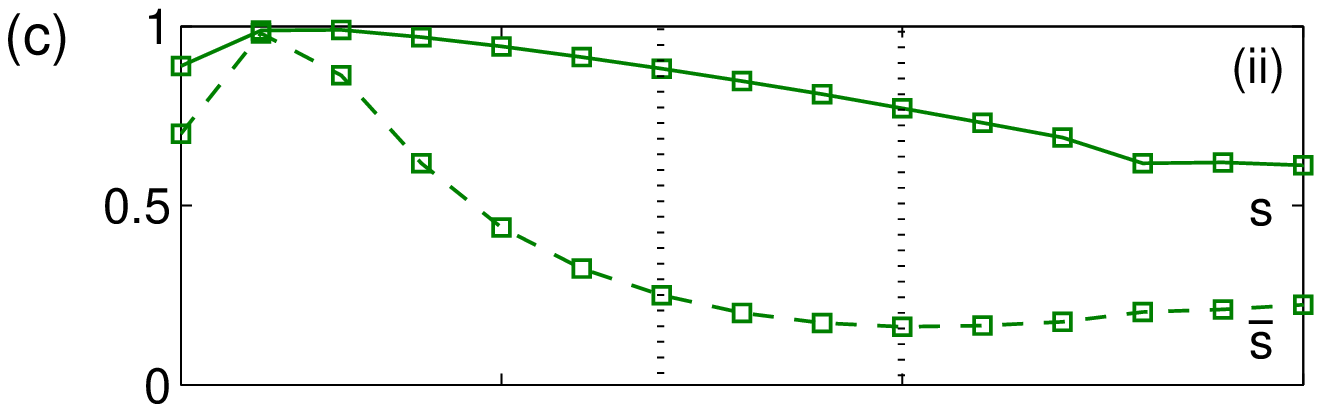} \\
\includegraphics*[width=9cm]{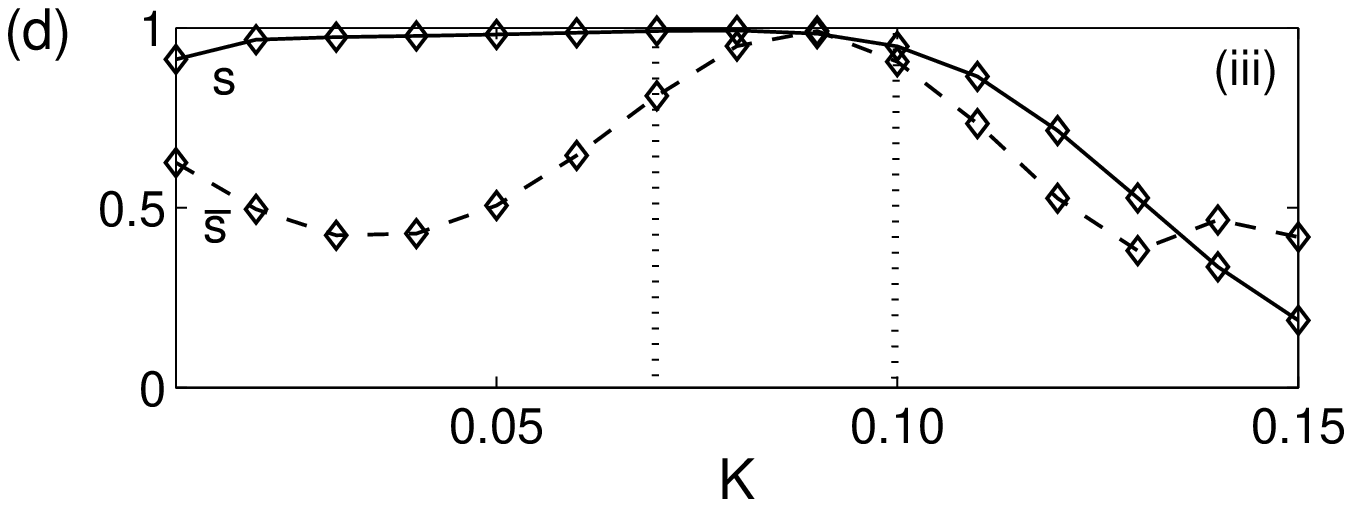}
\end{tabular}
\end{indented}
\caption{\label{holos} (a) The unitarity measure,
$s(\1,\Gamma_{C_l}\Gamma_{C_l}^\dagger)$, as a function of $K$ for the three
configurations given in Table \ref{parconfs}. The measure of
off-diagonality, $s(|R^2|,|\Gamma_{C_l}|)$ (\full), and the total
fidelity, $\bar{s}(R^2,\Gamma_{C_l})$ (\dashed), as a function of $K$
for the parametrizations (b) (i) (\opencircle), (c) (ii) (\opensquare) and (d) (iii)
(\opendiamond). Based on unitarity and the energy gap behavior, we expect a stable phase in the area $0.07 \lesssim K \lesssim 0.10$ bounded by the dashed vertical lines. }
\end{figure}

Since the spectrum varies slightly with $t$ during the braiding process, we
define the minimal fermion gap, $\Delta$, and the maximum energy splitting
between the two ground states, $\delta$, by
\begin{eqnarray}
\Delta = \min_{t} (\epsilon_3^t-\epsilon_2^t), \quad \delta =
\max_t (\epsilon_2^t - \epsilon_1^t), \label{gapdeg}
\end{eqnarray}
respectively, where $\epsilon^t_k$ is the $k$th eigenvalue at step $t$.
These are plotted in~\Fref{gaps}, where we observe that
both the fermion gap and the level of degeneracy improve as $K$ and $d$
increase. Under the adiabatic approximation the
holonomy corresponds to the exact time evolution when $\Delta \gg \delta$
and $\delta \rightarrow 0$. To physically accommodate these conditions in a finite
size system, the vortex transport should be fast enough compared to
$\delta$ for the states $\ket{\Psi_{\alpha}}$, $\alpha=1,2$ to appear as degenerate,
but slow enough compared to $\Delta$ so that no fermionic excitation is
produced. We see from~\Fref{gaps} that $\frac{\delta}{\Delta} \lesssim 10^{-2}$ for (iii) when $K \gtrsim 0.07$. This region can support the adiabaticity conditions and hence we take $K \approx 0.07$ as a lower bound for identifying a stable topological phase.

To quantitatively study the holonomy, we introduce a fidelity measure for a
target matrix $U$ and a test matrix $V$ as
\begin{eqnarray} \label{s}
    s(U,V) & = & \frac{1}{4} \tr \left( UV^\dagger + VU^\dagger \right).
\end{eqnarray}
For $U$, $V$ unitary $2\times2$ matrices we have that $s(U,V)=1$ if and only
if $U=V$, while in general $s(U,V)\leq1$. Let $\Gamma_{C_l}$ be the
numerically obtained holonomy with off-diagonal elements $re^{i\theta}$, $0
\leq r \leq 1$. After fixing the gauge\footnote{
The holonomy \rf{holonomy} is only given up to a gauge transformation $g:
\Gamma_C \to g \Gamma_C g^\dagger$ \cite{Pachos00}. Before $s(U,V)$ can be
evaluated, the gauge $g$ must be fixed. Due to the finite size of the system
the two ground states are never perfectly degenerate (see~\Fref{gaps}),
implying $g =\textrm{diag}(e^{i\phi_1},e^{i\phi_2})$ for some random phases $\phi_1$ and $\phi_2$. This can be easily
taken into account.}, we evaluate the unitarity
measure, $s(\1,\Gamma_{C_l}\Gamma_{C_l}^\dagger)$, \Fref{holos}(a), and
the two different fidelity measures of the holonomy:
$s(|R^2|,|\Gamma_{C_l}|)=r$ (measure of off-diagonality that characterizes
$R^2$) and $\bar{s}(R^2,\Gamma_{C_l})=\frac{1}{2}[s(R^2,\Gamma_{C_l})+1]
=\frac{1}{2}[r\cos(\frac{\pi}{4}+\theta)+1]$ (the total fidelity),
\Fref{holos}(b-d). Here $|U|$ denotes a matrix $U$ with its elements replaced by their absolute
values.

First, we observe that the unitarity
measure is above $98\%$ for all parametrizations when $K\lesssim0.10$, which we take as an upper bound for identifying a stable topological phase. For
(i) we obtain no significant off-diagonality due to the small size of the system.
However, for (ii) the holonomy is predominantly off-diagonal (e.g. $r >
0.9$) for $0.02\lesssim K\lesssim0.04$, and for (iii) for $0.02\lesssim
K\lesssim0.09$. The total fidelity, $\bar{s}$, accounts also for the overall
phase and can distinguish between the Ising ($\bar{s}=1$) and $SU(2)_2$
($\bar{s}=\frac{1}{2}$) anyon models whose monodromies only differ by an
overall phase factor, $e^{-i\pi/2}$ \cite{Rowell}. We observe that for (ii) there is a
small region around $K\approx 0.02$ and for (iii) there is a wider region,
$0.08\lesssim K\lesssim0.10$, where $\bar{s} >  0.9$. The maximum fidelities
are given by $0.981$ and $0.991$, respectively. Parametrization (iii)
also has a region $0.02\lesssim K\lesssim0.05$ where $\bar{s} \approx
\frac{1}{2}$ with error $\pm 10^{-1}$. However, we disregard this regime, because such a region does not exist for the smaller system (ii) and it lies outside the domain which we consider as a stable topological phase. Further, we check for all parametrizations and
all $K$ that $\Gamma_{C_{o}} \approx \1$ with error less than $10^{-2}$,
that for $K=0$ the holonomy vanishes and that $\Gamma_{C_{l}^{-1}}
=\Gamma_{C_{l}}^\dagger$ when the direction of braiding is reversed.

\section{Braiding and detection of the non-Abelian statistics in a laboratory}

Since our calculation involves only the experimentally accessible parameters $J$ and $K$, it translates directly
to how one could physically implement the creation and transport of anyons
in the laboratory. In particular, in the optical lattice proposal of Micheli et al. \cite{Micheli05}, the vortex transport would correspond, given sufficient site addressability, to the local adiabatic inversion of magnetic field as well as of the couplings $J$ through the introduction of suitably tuned lasers. Also, as the energy of the system is known to depend on the
zero mode populations~\cite{Lahtinen}, the effect of braiding can be
detected through spectral means. As the monodromy swaps these populations
between the vortex pairs, the energy behavior of the system will
be different when the vortices from a single pair are brought close together
before and after the braiding. Detecting this energy shift reveals the
non-Abelian statistics.

By evaluating the holonomy, we were able to identify a range of the model's parameters where the simulation approximates well the exact time evolution of a physical system and where the statistics corresponds to the Ising anyons. As expected, larger systems exhibit the predicted statistics with higher fidelity. The required magnitude of $K$, however, is larger than anticipated ($K\approx 0.1$ for parametrization (iii)). In the original work \cite{Kitaev05}, the three-body term appears in third order perturbation theory when one considers a general Zeeman term ($h \sum_\nu \sum_i \sigma_i^\nu$) as a perturbation. In our normalization the expansion is valid when $h^2 \ll 1$. Since $K \sim h^3$, for $K=0.1$ one can estimate $h^2 \approx 0.2$, which clearly does not satisfy the criteria. Therefore, in order to introduce the three-body terms into the Hamiltonian perturbatively, such as by adding a small magnetic field in the optical lattice proposal \cite{Micheli05}, one needs to consider larger systems. On the other hand, were the three-body terms engineered \cite{Buchler}, our calculation provides exact predictions for braiding experiments in such systems.

\section{Conclusions}

In summary, we formulated a method to directly study non-Abelian statistics in
exactly solvable lattice models whose ground state admits representation as a Slater determinant. By applying it to Kitaev's model, we
identified finite regions of the couplings, where the non-Abelian statistics
corresponds to Ising anyons. This confirms the previous conjectures for the
presence of a non-Abelian topological phase. Finally, we
proposed a scheme for the implementation and detection of non-Abelian
statistics in the laboratory. Such an experiment would be an important step
towards the physical realization of topological quantum computation. It is an interesting topic for future research to study whether the holonomy can be used as an order parameter for the topological phase when the system is subject to perturbations.

\section*{Acknowledgements}

We would like to thank Nick Read, Matthias Troyer and Zhenghan Wang for
inspiring conversations. This work is supported by EPSRC, the Finnish
Academy of Science, the EU Networks EMALI and SCALA and the Royal Society.

\end{document}